\documentclass[]{aa}
\usepackage{graphicx}
\usepackage{natbib}
\usepackage{amssymb}
\usepackage{amsmath}
\usepackage{xspace}
\usepackage{color}
\usepackage[varg]{txfonts}
\newcommand{\be}{\begin{equation}}
\newcommand{\ee}{\end{equation}}

\begin{document}

   \title{Transport of magnetic turbulence in Supernova remnants}

   \author{R. Brose\inst{1,2}\fnmsep\thanks{Corresponding author, \email{robert.brose@desy.de}}
          \and
	  I. Telezhinsky\inst{1,2}\fnmsep\thanks{Corresponding author, \email{igor.telezhinsky@desy.de}}
          \and
          M. Pohl\inst{1,2}}

\institute{DESY, 15738 Zeuthen, Germany 
\and Institute of Physics and Astronomy, University of Potsdam, 14476 Potsdam, Germany}

\date{Received ; accepted }

 
  \abstract
   {Supernova remnants are known as sources of galactic cosmic rays for their non-thermal emission of radio waves, X-rays, and gamma-rays. However, the observed soft broken power-law spectra are hard to reproduce within standard acceleration theory based on the assumption of Bohm diffusion and steady-state calculations.}  
   {We point out that a time-dependent treatment of the acceleration process together with a self-consistent treatment of the scattering turbulence amplification is necessary.}
  {We numerically solve the coupled system of transport equations for cosmic rays and isotropic Alfv\'enic turbulence. The equations are coupled through the growth rate of turbulence determined by the cosmic-ray gradient and the spatial diffusion coefficient of cosmic rays determined by the energy density of the turbulence. The system is solved on a co-moving expanding grid extending upstream for dozens of shock radii, allowing for the self-consistent study of cosmic-ray diffusion in the vicinity of their acceleration site. The transport equation for cosmic rays is solved in a test-particle approach.}
   {We demonstrate that the system is typically not in a steady state. In fact, even after several thousand years of evolution, no equilibrium situation is reached. The resulting time-dependent particle spectra strongly differ from those derived assuming a steady state and Bohm diffusion. Our results indicate that proper accounting for the evolution of the scattering turbulence and hence the particle diffusion coefficient is crucial for the formation of the observed soft spectra. {In any case, the need to continuously develop magnetic turbulence upstream of the shock introduces non-linearity in addition to that imposed by cosmic-ray feedback.}}
   {}


   \maketitle
%

\section{Introduction}

Diffusive shock acceleration (DSA) at the forward shock of a Supernova remnant (SNR) is an efficient process that relies on self-generated turbulence \citep{Blandford.1987}. Streaming cosmic rays (CRs) in the upstream region of the shock generate magnetic turbulence that enhances the acceleration process, which in turn leads to further turbulence growth. This process is terminated by escape of CRs or generally when the growth time of turbulence becomes longer than the evolutionary time scale of the system. Conventionally, it is considered that turbulence growth is the fastest process in the system, which is then followed by particle acceleration, and finally by global magneto-hydrodynamical (MHD) evolution of the SNR. It is then often assumed that for the first two processes a quasi-equilibrium develops which slowly changes on account of the MHD evolution. Under the assumption of steady state for both turbulence and particle transport, the cosmic-ray distribution at the shock can be derived accounting for their feedback \citep{2002APh....16..429B, Caprioli.2009}. It can then be imposed on global SNR models \citep{2012ApJ...744...39E}, or, alternatively, one can solve the entire coupled system of turbulence, CRs, and SNR fluid under steady-state conditions \citep{2014ApJ...789..137B}. It has been realized, however, that for SNR limited time is available both for turbulence growth and for particle acceleration \citep{1983A&A...125..249L, Bell.2013, Schure.2013}, and so the steady-state assumption for turbulence and CRs up to the highest energies is questionable.

Here, we introduce a fully time-dependent calculation of the cosmic-ray acceleration coupled to the evolution of isotropic Alfv\'enic turbulence using an analytical self-similar description of the SNR magnetohydrodynamics. {The advantage of our method lies in the full account of the time evolution of the hydrodynamical flow, the cosmic-ray distribution, and magnetic turbulence. Among the simplifications are the neglect of cosmic-ray feedback and the treatment of the shock as infinitely thin with parametrized cosmic-ray injection. The shock structure and particle pre-acceleration can be well studied with kinetic simulations, but even very large hybrid simulations \citep[e.g.][]{Caprioli.2014a} cover at most about one hour of real time and a region not larger than one Astronomical Unit. Steady-state Monte-Carlo studies including full feedback on the other hand may overestimate the available growth-time. {In \cite{2014ApJ...789..137B} about 180 years of optimal wave-growth would be needed to amplify the intensity of turbulence at the longest wavelengths ($k/k_0=10^{-6}$) by five orders of magnitude to its steady-state value, considering the intensity growth-time of 16 years. The simulation setup with a precursor size of $0.5\,$pc, which is reasonable for PeV-scale particles in a $40\,\mu$G field assuming Bohm-diffusion, provides only 100 years of time before the plasma has flown to the shock.} Our study is complementary to simulations of the small-scale physics operating at shocks and to steady-state Monte-Carlo studies including full feedback \citep{2014ApJ...789..137B} that neglect the evolution of the remnant. With our method we can examine the system over very long time, ten of thousands of years, and cover a considerable fraction of the lifetime of an SNR. Our model} accounts for competing plasma processes without the assumption of a steady-state situation. This renders it possible to study {additional} effects that arise from the limited time that is available for turbulence growth and particle acceleration. 

As a result we obtain the energy density of magnetic turbulence and particle spectra at any position on the grid, which allows to study the propagation of the CRs and their escape from the source.

\section{Particle acceleration}\label{sec:pa_ac}
To model cosmic-ray acceleration we use a kinetic approach in the test-particle approximation \citep{Telezhinsky.2012a,Telezhinsky.2012b,Telezhinsky.2013}. The feedback of accelerated CRs on the shock structure is negligible, as long as the CR pressure stays below 10\% of the shock ram pressure \citep{2010ApJ...721..886K}. Thus the acceleration process can be treated independent of the SNR evolution as long as the amount of energy contained in CRs is limited. In any case, the purpose of the present paper is to isolate and discuss the impact of time-dependent wave growth. 
We numerically solve the time-dependent transport equation for the differential number density of cosmic rays in spherically-symmetric geometry \citep{Skilling.1975a}:
\begin{align}
\frac{\partial N}{\partial t} &= \nabla(D_r\nabla N-\mathbf{u} N)-\frac{\partial}{\partial p}\left( (N\dot{p})-\frac{\nabla \cdot \mathbf{ u}}{3}Np\right)+Q
\label{CRTE}
\end{align}
where $N$ is the differential number density of cosmic rays, $D_r$ is the spatial diffusion coefficient, $\mathbf{u}$ is the advective velocity, $\dot{p}$ are the energy losses, and $Q$ is the source of thermal particles that is treated according to the thermal-leakage injection model \citep{Blasi.2005a}.
The transport equation is transformed to a frame co-moving with the shock in which the radial coordinate is normalized to the shock radius, $x=r/R_{sh}$. To resolve the precursor of lowest-energy CRs, we make another transformation, $(x-1) = (x^*-1)^3$. For equidistant binning in $x^*$ this transformation guarantees a very good resolution close to the shock while simultaneously extending out to several tens of shock radii in the upstream region, allowing for all injected particles to remain in the simulation domain.

\subsection{Diffusion coefficient}

One of the crucial but still poorly known parameters for the acceleration process and subsequent propagation is the spatial diffusion coefficient \citep{Yan.2012a, Telezhinsky.2012b}. It governs the efficiency of cosmic-ray acceleration and thus the maximum energy reached by the cosmic rays. It is also responsible for the spatial distribution of accelerated particles both upstream and downstream of the shock, that in turn impacts on the subsequent emission from the source and its vicinity. The diffusion coefficient is usually assumed to be Bohm-like, i.e.,
\begin{align}
 D_{r} = \frac{v}{3} r_g \eta \text{ , }
\label{diff_bohm}
\end{align}
where $v$ is the particle velocity, $r_g$ is its gyroradius, and $\eta$ is the ratio of background magnetic energy density to energy density in magnetic fluctuations and usually assumed to be order of unity for particles of all energies. The following arguments suggest that this approach is oversimplified. As was noted, the diffusion coefficient is directly connected to the magnetic-field fluctuations. Let us assume that CRs are being scattered by Alfv\'en waves that satisfy the resonance condition 
\begin{align}
k_\mathrm{res} = \frac{qB_0}{pc}\label{res_1}\text{ , }
\end{align}
where $k_\mathrm{res}$ is the wavenumber, $q$ is the particle charge, and $B_0$ is the background magnetic field. Then, the diffusion coefficient reads \citep{Bell.1978a,Blandford.1987}
\begin{align}
 D_{r} &= \frac{4 v}{3 \pi }r_g \frac{U_m}{\mathcal{E}_w}
\label{diff_1}
\end{align}
where $\mathcal{E}_w$ denotes the energy density per unit logarithmic bandwidth of Alfv\'en waves resonant with particles of momentum $p$ according to resonance condition (\ref{res_1}) and $U_m$ is the energy density of the background magnetic field $B_0$. Bohm diffusion (\ref{diff_bohm}), for which $\eta$ is a constant, is then equivalent to a featureless and flat magnetic turbulence spectrum {(see section \ref{sec:ener_dens})}.
But as pristine plasma is continuously advected toward the shock, magnetic turbulence must be constantly replenished in the upstream region. If the turbulence is a result of the growth of Alfv\'en waves (for instance due to resonant amplification by streaming CRs) as well as their spatial transport, compression at the shock, damping by various mechanisms, and spectral energy transfer due to cascading, it is obviously very unlikely that the turbulence spectrum in the SNR and its vicinity is flat and featureless. Besides, from $\gamma$-ray observations of SNRs and their surroundings we now understand that i) it is hard to accommodate Bohm diffusion for particles in an energy band as wide as we see in SNRs, and ii) the diffusion around SNRs is much slower than the average Galactic one, but much faster than Bohm. {This is in fact the expected behavior as turbulence must continuously be generated in the upstream region of the shock \citep{2011MNRAS.415.3434F,Yan.2012a}. A time-dependent calculation of the spectrum of magnetic turbulence is clearly needed to derive the self-consistent diffusion coefficient in the precursor region and should result in a more realistic and self-consistent picture of cosmic-ray acceleration in SNRs.}

\section{Magnetic turbulence}
\label{sec:turb_tran}

\subsection{Spectral energy density}
\label{sec:ener_dens}
We consider Alfv\'en waves as scattering centers for CRs. Alfv\'en waves can be considered a small contribution to the magnetic field at some position, so that 
\begin{align}
 B_\mathrm{tot} = B_0 + \delta b\text{ , }
\end{align}
where $\delta b$ is the combined amplitude of all waves present at the given position. Averaging the energy density over sufficiently large times gives 
\begin{align}
 B_\mathrm{tot}^2 = B_0^2+\langle \delta b^2\rangle\text{ . }
\end{align}
The total energy density in the waves can be represented as 
\begin{align}
\langle \delta b^2\rangle = 4\pi\int W_w (k)\, dk = 4\pi \int E_w (k)\, d\ln k \text{ , }
\end{align}
where $W_w$ and $E_w$ are the spectra of magnetic turbulence energy density per unit interval in $k$ and $\ln k$, respectively.

The energy density per unit logarithmic bandwidth of resonant Alfv\'en waves is then
\begin{align}
\mathcal{E}_w = \int E_w (k) \delta(k - k_\mathrm{res}) \, d\ln k \text{ , }
\end{align}
and the diffusion coefficient of a particle with momentum $p$ moving in the background field $B_0$ can be calculated using expression (\ref{diff_1}).

\subsection{Transport equation}

We consider a 1-D spherically-symmetric geometry and treat the turbulence as isotropic. To investigate the growth, damping, and cascading of the Alfv\'en waves, as well as their propagation along the background magnetic field, we solve an equation for the transport of the magnetic turbulence along with the transport equation (\ref{CRTE}) for cosmic rays. The transport of magnetic turbulence can be described by a continuity equation for the spectral energy density, $E_w = E_w(r,k,t)$:
\begin{align}
 \frac{\partial E_w}{\partial t} + \mathbf{u} \cdot (\nabla E_w) + C_{w}(\nabla \cdot \mathbf{u})E_w + k\frac{\partial}{\partial k}\left( {k^2} D_k \frac{\partial}{\partial k} \frac{E_w}{k^3}\right) = \nonumber\\
=2(\Gamma_g-\Gamma_d)E_w
\label{Turb_1}
\text{ , }
\end{align}
where $C_{w}$=1.5 denotes the pre-factor for wave compression at the shock \citep{McKenzie.1982}, $D_k$ is the diffusion coefficient in wavenumber space representing cascading, and $\Gamma_g$ and $\Gamma_d$ are the growth and the damping rates, respectively.
Therefore, we consider growth, damping, advection, compression of turbulence at the shock, as well as spectral energy transfer through cascading. The transport equation for magnetic turbulence (\ref{Turb_1}) is transformed to a frame co-moving with the shock, in the same manner as the cosmic-ray transport equation (\ref{CRTE}), leading to a system of two coupled equations which we solve numerically using implicit finite-difference methods \citep{FiPy.2009a}.

\subsection{Wave growth}

Particles streaming faster than the Alfv\'en speed should generate Alfv\'en waves at wavelengths similar to the gyroradii of the particles (\citet{Wentzel.1974} and references therein). In the diffusion limit the growth rate of waves can be related to the pressure gradient of the CRs. This mechanism is known as resonant amplification of Alfv\'en waves and the only wave-driving process considered in this paper. The growth rate due to resonant amplification is given as \citep{Skilling.1975a, Bell.1978a}
\begin{align}
 \Gamma_{g} &= \frac{v_A p^2v}{3E_w}\left|\frac{\partial N}{\partial r}\right| \label{Bell_res}\text{ . }
\end{align}
Magnetic turbulence can also be produced on very small scales through Bell's non-resonant instability \citep{2000MNRAS.314...65L, 2004MNRAS.353..550B}. The interaction of cosmic rays with this mode is also nonresonant and does not involve pure pitch-angle scattering \citep{1984JGR....89.2673W,2010ApJ...709.1148N}. Presumably, the mean free path for scattering is small only for very low-energy particles whose Larmor radius is commensurate the wavelength of Bell's mode \citep{2014ApJ...789..137B}. Those particles are typically not present far out in the cosmic-ray precursor, and so the impact of Bell's mode is likely moderate {at later times. Generally Bell's mode might be operating only for a few e-foldings. Following \cite{2008ApJ...684.1174N}, assuming Bohm-diffusion and comparing the growth-time to the shock-capture time, one finds for the available number of e-foldings N, 
\begin{align}
 N &= 3.3 \left( \frac{v_{sh}}{4000\,\text{km/s}}\right)\left( \frac{v_{A}}{10\,\text{km/s}}\right)^{-1}\left( \frac{U_{CR}}{0.1U_{bulk}}\right)\,,
\end{align}
where $U_{CR}$ are the energy-density of the cosmic rays and $U_{bulk}$ the energy density of the bulk plasma respectively. In our fully time-dependent treatment another non-linearity would occur, as initially only low-energetic particles are present in the simulation, whose current is quickly reduced by a decreasing diffusion-coefficient on account of wave-growth. The fastest growth-rate would initially occur on scales where Ion-cyclotron damping plays a role and cascading is fast. The resulting total growth-rate $\Gamma_g-\Gamma_d$ would thus be smaller.}

{Recently, fast-mode waves were found to be efficient scatterers of cosmic rays through both resonant and nonresonant interactions \citep{2004ApJ...614..757Y}. Compressive modes such as fast-mode waves are thermally damped with a rate that depends on the orientation of wavevector, and so a 3D treatment of the wave spectrum would be needed which we defer to a future publication. Likewise, we ignore large-scale modes that are driven by the cosmic-ray pressure or are produced as response to small-scale turbulence \citep{2009ApJ...707.1541B,2011MNRAS.410...39B,2011MNRAS.418..782S}. }

\subsection{Wave damping}

For wave damping we consider neutral-charged collisions and ion-cyclotron damping. The damping rate due to neutral-charged collisions is given as \citep{Kulsrud.1971a, Bell.1978a}
\begin{align}
 \Gamma_{d,nc} = \frac{1}{2}n_H\langle v\sigma\rangle
\end{align}
where $n_H$ is the number density of neutral hydrogen and $\langle v\sigma\rangle$ is the 
velocity-weighted cross section, averaged of the random velocity of ions.
Normally, neutral-charged damping is relatively weak and independent of the wavenumber of Alfv\'en waves. This mechanism is mostly important in regions of low temperatures and high densities such as molecular clouds. As was noted before, if cosmic rays penetrate molecular clouds, their spectrum may be strongly modified due to evanescence of magnetic turbulence that is responsible for particle scattering \citep{Malkov.2011a}. In this work the effect of the neutral-charged damping is negligible since no molecular clouds or other regions of sufficient low temperature and ionization fraction are considered.

Ion-cyclotron damping is due to interaction of Alfv\'en waves with the thermal particles of the plasma and is strongest at small scales 
\begin{align}
 \Gamma_{d,IC} = \frac{v_A c k^2}{2 \omega_P}\text{ , }
\end{align}
where $\omega_P$ is the ion plasma frequency \citep{Threlfall.2011a}. This damping should transfer energy to the plasma via heating, which is not yet considered in this work, though we are aware that it might modify the spectrum around the scale of particle injection.

\subsection{Wave cascading}

The process of energy transfer through cascading from larger scales to smaller scales is not yet fully understood and subject of active research. Empirically it can be described as a diffusion process in wavenumber space \citep{Zhou.1990, Schlickeiser.2002a}. Given the assumption of isotropic turbulence, the corresponding diffusion coefficient is 
\begin{align}
 D_k = k^{3} v_A\sqrt{\frac{E_w}{2 B_0^2}}\text{ . }
\end{align}
If cascading is the dominant process, this phenomenological treatment will result in a Kolmogorov-like turbulence spectrum, $E_w\propto k^{-2/3}$.
Because cascading is treated as a diffusion process, a small fraction of energy is transferred also to scales larger than the turbulence injection scale. This permits scattering of particles whose energy is higher than those currently driving the turbulence and is in effect similar to resonance broadening. The acceleration time for these particles is therefore reduced and the acceleration process more efficient.

\subsection{Initial conditions}
In the absence of initial turbulence there is nothing to grow, and so Eq.~(\ref{Turb_1}) requires some seed turbulence. We therefore take as an initial condition an ISM turbulence derived with Eq.~(\ref{diff_1}) from the diffusion coefficient, 
\begin{align}
 D_0 = 10^{27}\left(\frac{pc}{10\,\text{GeV}}\right)^{1/3}\left(\frac{B_0}{3\,\mu\text{G}}\right)^{-1/3}\label{D_r} \text{ . }
\end{align}
The value of $D_0$ is a factor $100$ lower than the ISM diffusion coefficient found in studies of galactic CR propagation \citep[e.g.]{2011ApJ...729..106T}, {corresponding to a somewhat enhanced  intensity of turbulence at large distance from the forward shock. This choice is computationally expedient and increases numerical stability during the initial stages of our simulations. As long as the energy density of the turbulence attained through self-consistent amplification is orders of magnitudes higher than the initial intensity, the diffusion of particles near the shock is entirely governed by the self-generated turbulence. The intensity of turbulence is either determined by balance of growth and damping or cascading, in which case the intensity is insensitive to its initial value, or by the available time. In the latter situation our elevated initial intensity makes it \emph{easier} to reach a very high level of turbulence. Any failure to generate the turbulence intensity needed for very efficient cosmic-ray acceleration  is therefore intrinsic.}

Moreover, though not particularly important for the scope of the current paper, this diffusion coefficient allows all injected particles to stay within the numerical grid.

\section{Evolution of Supernova remnant}

The evolution of an SNR can be subdivided into three major stages: free expansion, adiabatic phase, and radiative stage. The initial free expansion is characterized by the fastest shocks, and cosmic-ray acceleration should be most efficient, but this stage is brief for a typical type-Ia SNR. Already after several hundreds years the shock sweeps up enough ISM to slow down considerably. The following adiabatic phase, given its long duration, should be most frequent among Galactic type-Ia SNRs. Because the shock is still rather fast, the bulk of the CRs produced by the remnant should be accelerated during this stage. In the subsequent radiative stage, the shock becomes very slow, allowing for efficient recombination in the downstream region. The layer of cooling material just behind the radiative shock contains a high fraction of neutrals, and so acceleration should level off.

\begin{figure*}[!ht]
\includegraphics[width=0.99\textwidth]{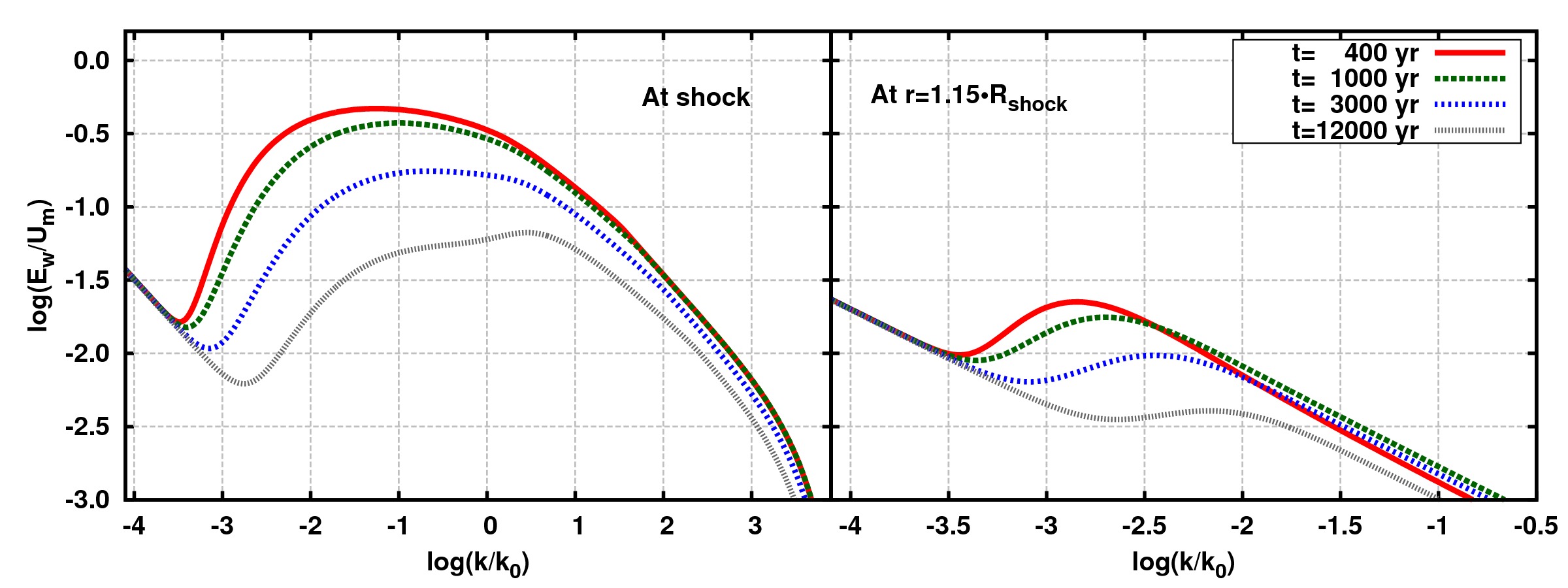}
\includegraphics[width=0.99\textwidth]{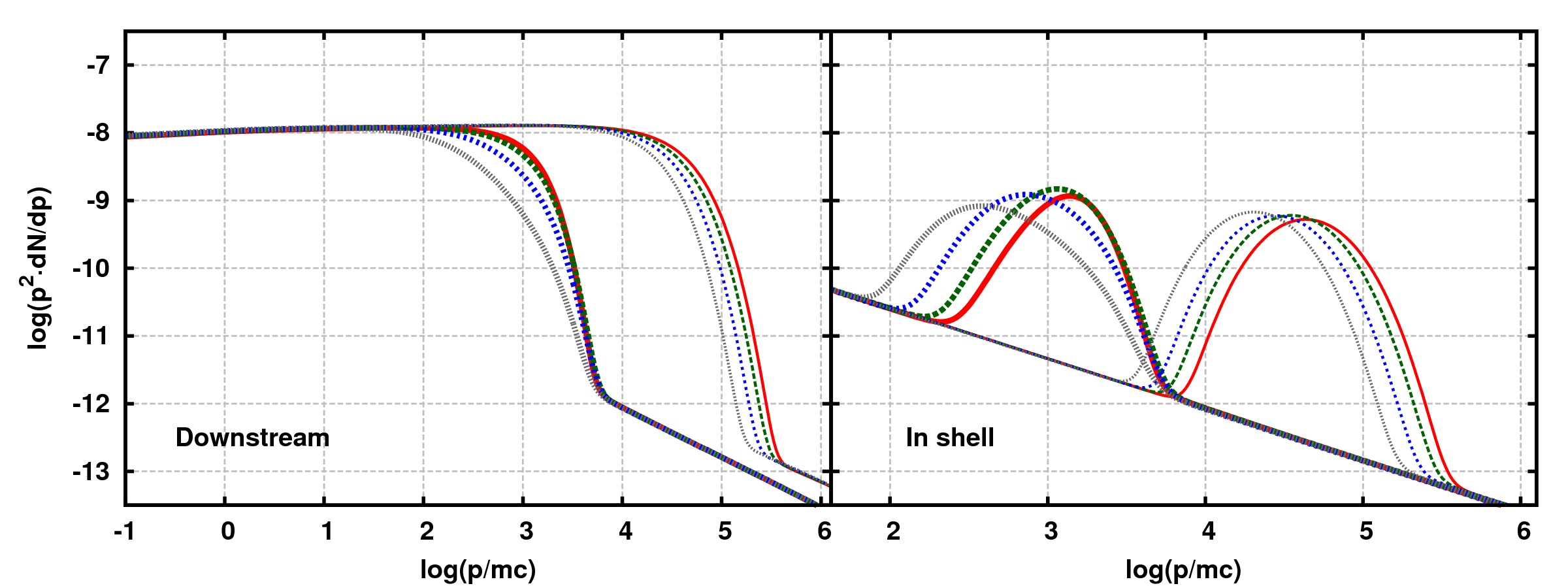}
\caption{Spectral evolution of turbulence energy density, $E_w$ (top), and differential proton number density, $N$ (bottom), for an SNR in the adiabatic stage at the age of 400 (red), 1000 (green), 3000 (blue) and 12000 (grey) years. Both the self-consistent treatment (thick lines) and Bohm-like diffusion (thin lines) are presented. Spectra for the location at $r=1.15R_{shock}$ are averages over a shell with thickness $0.5\,$pc. Particles with kinetic energy of 1 GeV are resonant with waves of wavenumber $k_0$.}
\label{fig:SedTu}
\end{figure*}
For core-collapse SNRs evolving in the winds of progenitor stars the picture is completely different. On account of strong variation in the wind parameters, the sequence of stages may be mixed up. Most of the core-collapse SNRs require a thorough numerical modeling, while type-Ia SNRs can be well described analytically. Besides, it was shown \citep{Telezhinsky.2015} that type-Ia SNRs should be easiest to observe with the future CTA facility owing to a continuous increase of $\gamma$-ray flux and size. For many core-collapse SNRs detectability and resolvability is a strong function of the age.

As here we aim at introducing our method, we consider an SNR in the adiabatic (Sedov-Taylor) stage evolving in a typical ISM with density of $0.4$~cm$^{-3}$ and an magnetic-field strength ($B_0$) of $5$~$\mu$G. We use analytic approximations for Sedov-Taylor solutions \citep{Cox.1981} to describe the evolution of the plasma-flow parameters. The corresponding self-similar solutions for the background magnetic field profiles inside the SNR are taken from \citet{Korobeinikov.1964}.

For the sake of comparison, we also performed simulations for two cases of freely expanding SNR in a uniform and power-law density medium. The simulations were carried out just for the first 1000 years. For these simulations we used analytic expressions for the shock evolution \citep{Truelove.1999a}, and the background magnetic field was calculated by solving the induction equation as described in \citet{Telezhinsky.2013}.

\section{Results}\label{sec:res}

We explore the evolution of magnetic-turbulence spectra and the corresponding particle spectra in adiabatic SNRs up to an age of 12000 years. We also present and discuss the results for free-expansion solutions. We provide a comparison with Sedov SNRs of the same ages. The results obtained from self-consistent calculations are then compared to the standard approach assuming Bohm diffusion. The specific model features Bohm diffusion everywhere downstream of the shock, whereas in the upstream region an exponential transition to the Galactic diffusion is assumed to occur between the forward shock and a location one SNR radius ahead \citep{Telezhinsky.2012b, Telezhinsky.2013}. In all calculations we use the same injection parameters. In order not to violate test particle approximation, we set the injection of thermal particles so that the cosmic-ray pressure always remains below 10\% of the shock ram pressure.

\subsection{Sedov-Taylor stage} 

\subsubsection{Turbulence spectra}

The evolution of magnetic turbulence is given at Fig.~\ref{fig:SedTu}. The turbulence spectrum corresponding to Bohm diffusion in these coordinates, as can be seen from Eq.~\ref{diff_1}, would be a constant line with a value of 4/$\pi$. In contrast, our calculations show that the turbulence spectra at the shock exhibit a complicated shape; there is a very extended region of efficient growth that spans over several orders of magnitude in $k$-space. This is not surprising, because at the position of the shock the turbulence is driven by particles of all energies, and, since low-energy particles dominate the cosmic-ray spectrum, the growth of turbulence is fastest at large $k$. However, the larger $k$ is, the more important cascading becomes, and so at some point it starts playing a crucial role and dominates over growth. This is a region where a transition to the inertial range happens, and a break in the spectrum appears. Finally, at high $k$, a classical Kolmogorov power-law turbulence spectrum is observed. An interesting result of such an interplay between growth and cascading is that in a rather wide range of $k$ the turbulence spectrum appears to be plateau-like and similar to that for Bohm diffusion.

\begin{figure*}[!ht]
\includegraphics[width=0.99\textwidth]{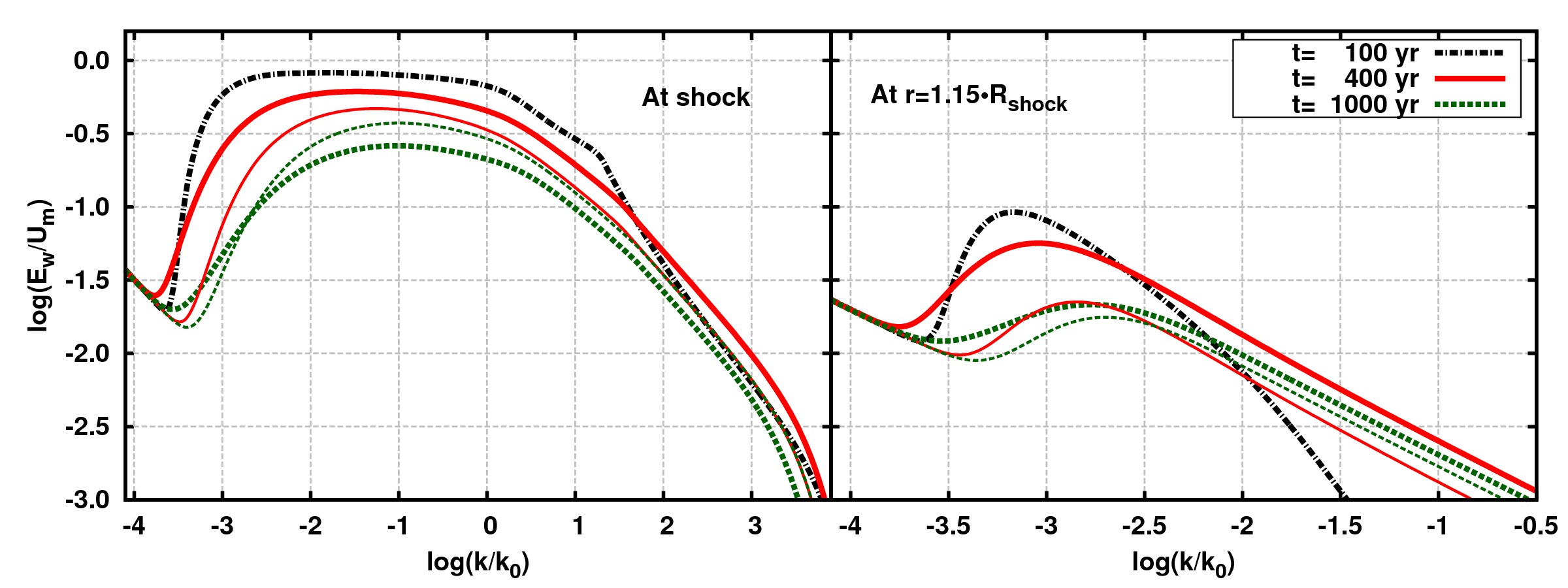}
\includegraphics[width=0.99\textwidth]{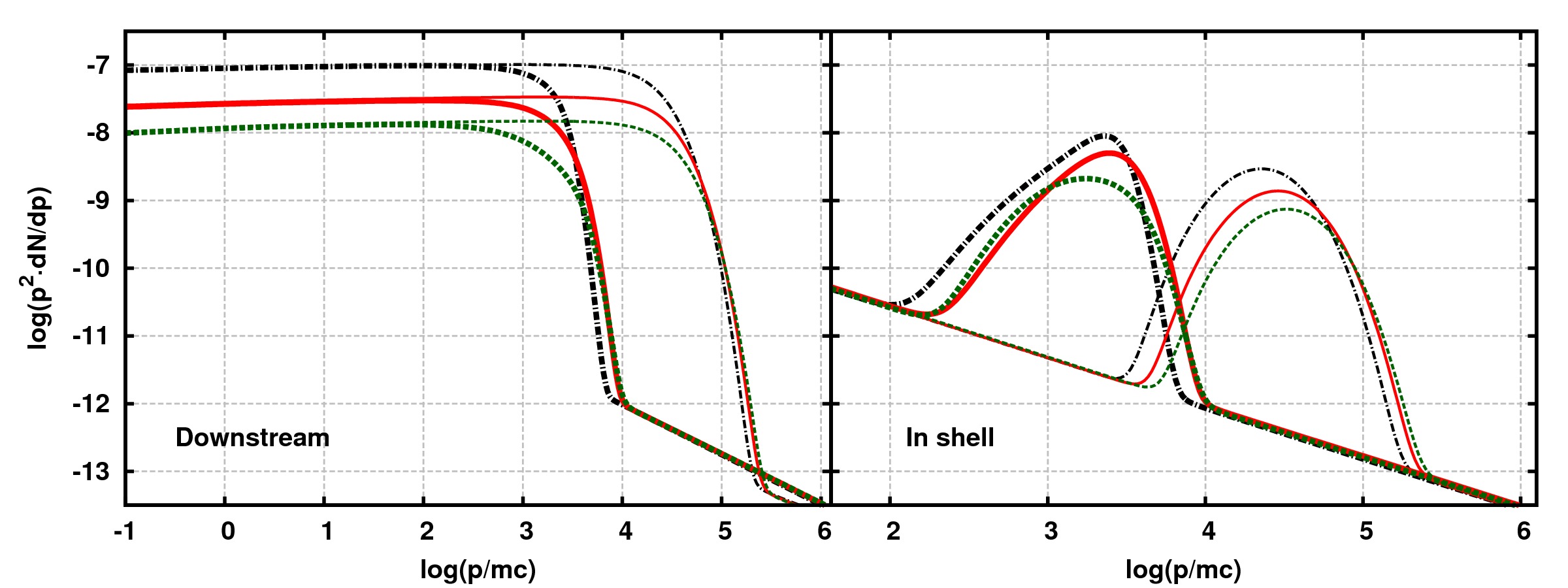}
\caption{Spectral evolution of turbulence energy density, $E_w$ (top), and differential proton number density, $N$ (bottom), for an SNR freely expanding into a uniform ISM (thick lines) at the age of 100 (black), 400 (red), and 1000 (green) years. For comparison, $E_w$ and $N$ derived with Bohm-like diffusion are given with thin lines. The shell and $k_0$ are the same as at Fig.~\ref{fig:SedTu}.}
\label{fig:ExpUni}
\end{figure*}
This interpretation is well supported by the shape of the spectrum at some distance upstream. The turbulence spectrum at the shock of a young SNR suggests that particles in a wide energy band should be well confined to the shock. However, at low $k$ there is a sharp cutoff in the turbulence spectrum, and thus particles beyond some energy can freely escape from the shock region. Therefore, at some distance from the shock in the upstream region one expects to see a narrow particle distribution that is peaked at the energy of escaping particles. This particle distribution should drive turbulence only in a very narrow k-region. For instance, at a location 15\% of the SNR radius ahead of the shock, there are too few low-energy particles to substantially impact on the turbulence growth, whereas high-energy cosmic rays can easily diffuse further away from the shock. Hence the turbulence spectrum looks like a classical one (Fig.~\ref{fig:SedTu}, right): we observe just two regions -- the first is the injection region peaked at the wavenumber of maximum growth corresponding to the peak in the particle energy spectrum, and the second is a cascading-dominated region at higher wavenumbers. The damping range, that would be the third region of the classical turbulence spectrum, is not shown at Fig.~\ref{fig:SedTu} in the upstream region because it is not important for this work.

With passage of time, the sharp cutoffs in the turbulence spectra at the shock become mild and the peaks in the upstream spectra broader. This can be explained by the initially high growth rates and rather strong magnetic-field amplification, so that only particles at the highest energy were escaping the shock. Later on, the cosmic-ray gradients are not as sharp, and the growth rates become low. Consequently, the turbulence level drops, allowing escape of particles over a wide range in energy. Additionally, the advection of turbulence from far upstream broadens the observed spectral distribution at later times.

To be noted from both plots is that there is no steady state. The spectral shape and the maximum level of turbulence in both the shock and the upstream regions are continuously changing.

\subsubsection{Particle spectra}

To understand the impact of the self-consistent, fully time-dependent coupled treatment of magnetic turbulence and cosmic-ray acceleration, we compare the particle spectra obtained here to those derived earlier assuming Bohm diffusion. The downstream volume-integrated spectra for that case are shown at the bottom panel of Fig.~\ref{fig:SedTu}.


Both calculations show similar power-law indices, but the cutoff regions differ sharply. In the case of Bohm diffusion, a classical exponential cutoff is seen at very high energy. In contrast, the self-consistent calculation yields systematically lower maximum energies and a smoother cutoff that is softer than exponential. Moreover, the spectral softening in the cutoff region shows an evolutionary trend, that arises because the diminishing turbulence amplitude provides for more effective particle escape as time passes. Escape becomes possible for particles in a broad range of energies, and the maximum particle energy decreases faster than for Bohm diffusion. In the Bohmian case of a constant turbulence amplitude, cosmic-ray escape is possible only at the high-energy tail of the distribution. 

The evolution of the maximum energy of particles and the history of their escape from the shock is best illustrated in the distribution of the particles upstream of the shock. Let us consider a particle population in a shell upstream at some distance from the shock. Ahead of the shock, it is expected that the particle distribution is dominated by escaping particles. For Bohm diffusion this is a log-parabola \citep{Bykov.2011a} centered around the maximum energy, exactly what we observe for the Bohm case at Fig.~\ref{fig:SedTu} (bottom right). The contrast to the self-consistently calculated spectra is huge, which resemble a log-parabola only at the very beginning when the turbulence amplitude is sufficiently high. (Fig.~\ref{fig:SedTu}, right). Later on, when the turbulence is weaker, the distribution of escaped particle shifts to lower energy and becomes substantially broader. One consequence is a softening and broadening of the cutoff region of the downstream particle spectrum, which under certain conditions can be fitted with a power-law and a cutoff. Steady-state simulations that incorporate the CR-feedback on the shock do not show this softening. Instead a hardening at the highest energies is observed \citep{2014ApJ...789..137B}. The non-linear impact of particles at the highest energy is the slowest process in the system, though, and it {is possible that the limited growth-time for waves scattering particles of the highest energies might attenuate the development of convex spectra and the spectral hardening at the highest energies typical for NDSA. We note that the hybrid-simulations of \cite{Caprioli.2014a}, that include the nonlinear contributions, do not show any deviation from the DSA-predicted ${s=-2}$ spectra, despite a transfer of 10\%-20\% of the total energy to accelerated particles. In the end, the need to continuously develop magnetic turbulence upstream of the shock introduces non-linearity in addition to that imposed by cosmic-ray feedback.} 

\begin{figure*}[!ht]
\includegraphics[width=0.99\textwidth]{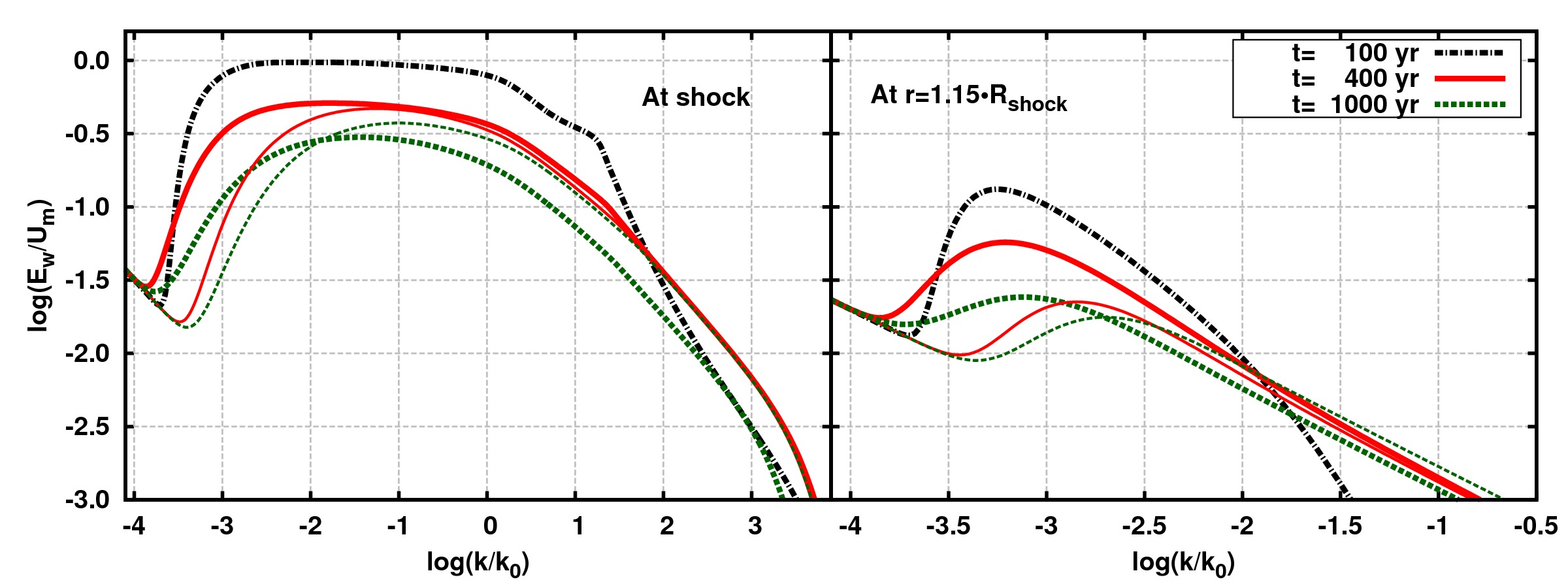}
\includegraphics[width=0.99\textwidth]{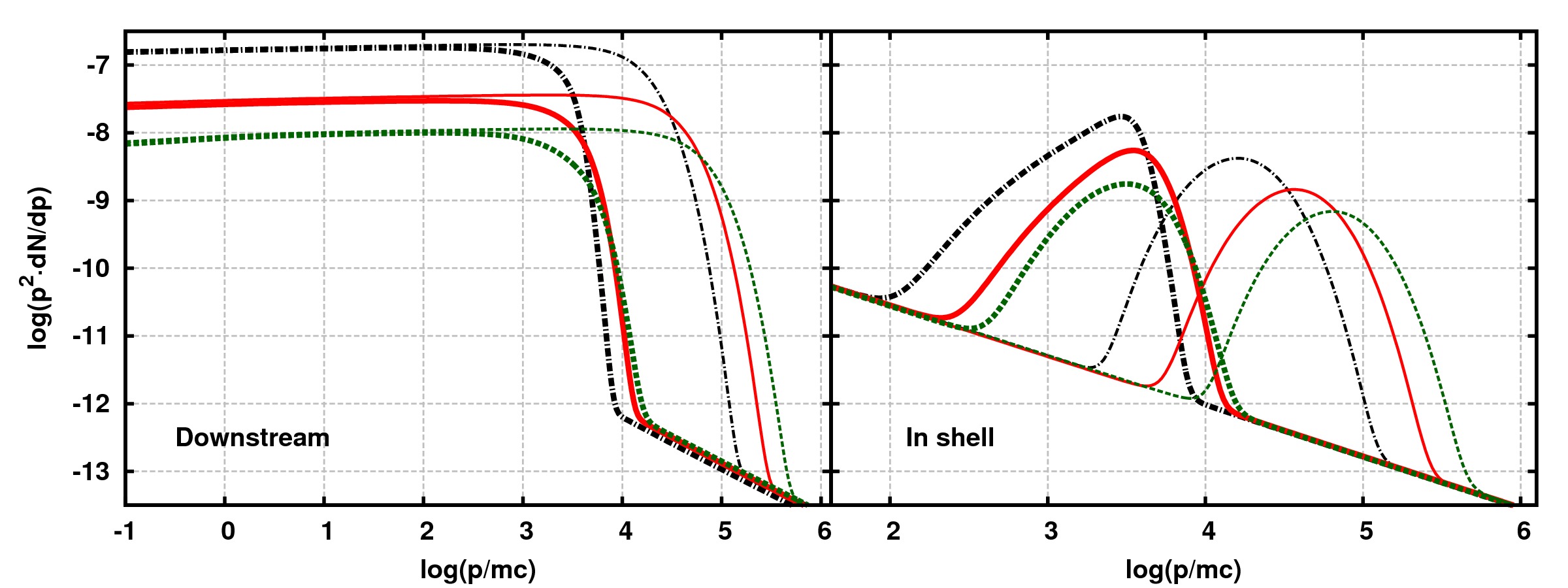}
\caption{Same as Fig.~\ref{fig:ExpUni} but for an SNR expanding into a wind zone.}
\label{fig:ExpWind}
\end{figure*}

In general, the time-dependent treatment introduces a connection between the maximum energy and the injection parameter. To reach higher energies more particles have to be injected because the growth rate (\ref{Bell_res}) used here is proportional to the gradient of the CR-distribution. This gradient determines the maximum level of turbulence which defines the cutoff energy. Moreover, the injection parameter is no longer an arbitrary parameter that simply scales the CR number density and thus photon fluxes. As in calculations of non-linear diffusive shock acceleration with cosmic-ray feedback \citep[e.g.][]{2006MNRAS.371.1251A}, it can be possibly constrained by accommodating both flux intensity and cutoff energy in the observational data. 

The results for the Sedov-Taylor stage are somewhat affected by our using a test-particle approach. Keeping the cosmic-ray pressure below 10\% of the ram pressure at late times requires a low injection efficiency, $\psi$, resulting in a maximum particle energy of only about $1\,$TeV. Although we did not calculate electron spectra and their emission here, the cutoff of the turbulence spectrum in our simulation does not allow for the high electron energies needed to explain observations of non-thermal X-ray emission from the remnants like SN1006 \citep{Koyama.1995a}. Our model would still be applicable to young remnants, where higher $\psi$-values can be used. It also indicates that there is a need for further improvements in the injection model. Arguably, the combination of particle feedback and a fully time-dependent treatment would be highly desirable.


\subsection{Free expansion stage}

We simulated two cases for the free-expansion stage of the SNR: i) a shock expanding in a uniform medium of constant density, and ii) a shock expanding into a wind-blown bubble. We terminate our calculations when the age reaches 1000 years.

\subsubsection{Turbulence spectra}

The turbulence spectra for the self-consistent simulations of the freely expanding SNR are shown at Fig.~\ref{fig:ExpUni} and Fig.~\ref{fig:ExpWind} for a uniform ambient medium and a wind zone with density $\rho\propto r^{-2}$, respectively. For both free-expansion runs, the maximum level of turbulence is higher, and spectra extend to lower $k$ compared to Sedov-Taylor cases of the same age. This is a consequence of the slower shock deceleration than in Sedov-Taylor solutions. Therefore, there are more particles of higher energy that can drive turbulence at small $k$. Other than that, the spectra are similar to those for the Sedov-Taylor phase.

There are slight differences in spectral evolution between the two free-expansion simulations. Again they mainly arise from the difference in the evolution of the shock velocity. In the wind zone, the shock speed does not significantly decrease during the simulation. Hence, spectra at the shock and in the upstream region tend to extend to lower $k$ than they do in uniform-medium simulations, and with time this difference becomes more pronounced. 

Besides, in wind-zone simulations the turbulence spectra in the upstream region have initially a somewhat concave structure beyond the peak. This can be seen comparing the red lines (400$\,$years) in the plot for the upstream-turbulence spectra (Upper right panels) for a uniform ambient medium (Figure~\ref{fig:ExpUni}) and for expansion into a wind zone (Figure~\ref{fig:ExpWind}). In the case without a windzone, the turbulence spectra for our time-dependent treatment and for the Bohm assumption are parallel, both displaying a Kolmogorov-like $k^{-2/3}$ scaling for $\log(k)>-3.25$. In the case with a windzone, the self-consistently calculated turbulence spectrum is clearly softer than that corresponding to Bohm diffusion, at least {for $\log(k)<-1.5$, on account of the limited time available for turbulence cascading in the presence of a very young, fast shock and the absence of low-energetic particles that could amplify turbulence.}

\subsubsection{Particle spectra}

The particle spectra shown at Fig.~\ref{fig:ExpUni} and Fig.~\ref{fig:ExpWind} for the uniform and wind-zone scenario, respectively, resemble those obtained with Bohm-like diffusion. The maximum energies are consistently lower, though, at least for our choice of a low injection efficiency. Also, the maximum energy does not increase as it would for Bohm-like diffusion, but stays fairly constant (see distributions of escaped particles at Fig.~\ref{fig:ExpUni} and Fig.~\ref{fig:ExpWind}). The slowly changing velocity during the free-expansion stage is compensated by the evolution of the turbulence spectra, and we do not observe substantial growth in maximum energy. There is no spectral softening observed as was seen in Sedov-Taylor solutions. The simulation time of 1000~years is too short for a decreasing turbulence level to have an significant effect on particles escape from the remnant, and the distributions of escaped cosmic rays look rather narrow and do not strongly deviate from a log-parabola.



\section{Conclusions}\label{sec:concl}

We developed a model for particle acceleration in SNR by simultaneously solving time-dependent transport equations for magnetic turbulence and cosmic rays. The equations are solved in spherically-symmetric geometry but so far are limited to the test-particle regime. We consider the cosmic rays being scattered by isotropic Alfv\'enic turbulence that is subject to compression, advection, cascading, damping and growth due to resonant amplification of Alfv\'en waves.
The calculated turbulence spectra, independent on the hydrodynamical model, reveal the same features at the shock: 
\begin{itemize}
\item a sharp cutoff at low $k$ numbers
\item a wide plateau-like region at intermediate $k$ that resembles that leading to Bohm-like diffusion 
\item a cascade-dominated tail at high $k$, where the spectrum is Kolmogorov-like
\item a damping-dominated part at very high $k$.
\end{itemize}
A wide plateau-like region in the $k$-spectrum is essential for particle acceleration.

We found that even for old remnants a steady state {is not reached, neither in turbulence nor in particle spectra. Even after 12000 years of SNR evolution both spectra are continuously changing,} which raises concerns as to the validity of cosmic-ray acceleration models that rely on the assumption of a steady state, either openly or implicitly. In particular, the spectral shape and intensity of turbulence are strongly time dependent. As a feedback, the particle spectra acquire significant features, which do not appear for Bohm-like diffusion. The more efficient escape in the self-consistent treatment gives rise to the formation of softer spectra at late stages of SNR evolution. {We note with interest that soft spectra, without any sign of hardening at the highest energies as predicted by NDSA-models, are indeed observed in the high-energy gamma-ray band \citep{Fermi.2013a}.} The need to continuously develop magnetic turbulence upstream of the shock introduces non-linearity in addition to that imposed by cosmic-ray feedback. Although our choice of a low injection efficiency is partly responsible for the low maximum energy of cosmic rays compared to that in the Bohm case, the evolutionary trend in maximum energy suggests that its evolutionary decrease proceeds much faster, and therefore at the age of $\gamma$-ray emitting galactic SNR the maximum energy is indeed lower than is estimated with steady-state models.


\begin{acknowledgements}
We acknowledge support by the Helmholtz Alliance for Astroparticle Physics HAP funded by the Initiative and Networking Fund of the Helmholtz Association.
\end{acknowledgements}


\bibliographystyle{aa}
\bibliography{References}








\end{document}